\documentclass[journal=ancac3,manuscript=article]{achemso} \usepackage{graphicx}
\usepackage{epstopdf}
\usepackage{amsmath}
\usepackage{color} \DeclareGraphicsExtensions{.pdf,.eps,.png,.jpg,.mps}

\author{Xiao-Fen Qiao}
\altaffiliation{Contributed equally to this work} \affiliation{State Key Laboratory of Superlattices and Microstructures, Institute of Semiconductors, Chinese Academy of Sciences, Beijing 100083, China}

\author{Jiang-Bin Wu}
\altaffiliation{Contributed equally to this work} \affiliation{State Key Laboratory of Superlattices and Microstructures, Institute of Semiconductors, Chinese Academy of Sciences, Beijing 100083, China}

\author{Lin-Wei Zhou}
\altaffiliation{Contributed equally to this work} \affiliation{Department of Physics, Renmin University of China, Beijing 100872, China}

\author{Jing-Si Qiao}
\affiliation{Department of Physics, Renmin University of China, Beijing 100872, China}

\author{Wei Shi}

\affiliation{State Key Laboratory of Superlattices and Microstructures, Institute of Semiconductors, Chinese Academy of Sciences, Beijing 100083, China}

\author{Tao Chen}
\affiliation{State Key Laboratory of Superlattices and Microstructures, Institute of Semiconductors, Chinese Academy of Sciences, Beijing 100083, China}

\author{Xin Zhang}
\affiliation{State Key Laboratory of Superlattices and Microstructures, Institute of Semiconductors, Chinese Academy of Sciences, Beijing 100083, China}

\author{Jun Zhang}
\affiliation{State Key Laboratory of Superlattices and Microstructures, Institute of Semiconductors, Chinese Academy of Sciences, Beijing 100083, China}

\author{Wei Ji}
\email{wji@ruc.edu.cn}
\affiliation{Department of Physics, Renmin University of China, Beijing 100872, China}

\author{Ping-Heng Tan}

\email{phtan@semi.ac.cn}
\affiliation{State Key Laboratory of Superlattices and Microstructures, Institute of Semiconductors, Chinese Academy of Sciences, Beijing 100083, China}

\title{Polytypism and Unexpected Strong Interlayer Coupling of two-Dimensional Layered ReS$_2$}

\begin{document}

\begin{abstract}
The anisotropic two-dimensional (2D) van der Waals (vdW) layered materials, with both scientific interest and potential application, have one more dimension to tune the properties than the isotropic 2D materials. The interlayer vdW coupling determines the properties of 2D multi-layer materials by varying stacking orders. As an important representative anisotropic 2D materials, multilayer rhenium disulfide (ReS$_2$) was expected to be random stacking and lack of interlayer coupling. Here, we demonstrate two stable stacking orders (aa and $\bar{a}b$) of N layer (NL, N$>$1) ReS$_2$ from ultralow-frequency and high-frequency Raman spectroscopy, photoluminescence spectroscopy and first-principles density functional theory calculation. Two interlayer shear modes are observed in aa-stacked NL-ReS$_2$ while only one interlayer shear mode appears in $\bar{a}b$-stacked NL-ReS$_2$, suggesting anisotropic-like and isotropic-like stacking orders in aa- and $\bar{a}b$-stacked NL-ReS$_2$, respectively. The frequency of the interlayer shear and breathing modes reveals unexpected strong interlayer coupling in aa- and $\bar{a}b$-NL-ReS$_2$, the force constants of which are 55-90\% to those of multilayer MoS$_2$. The observation of strong interlayer coupling and polytypism in multi-layer ReS$_2$ stimulate future studies on the structure, electronic and optical properties of other 2D anisotropic materials. \end{abstract}

Two dimensional (2D) van der Waals (vdW) layered materials have attracted great interest, due to its remarkable physical properties and excellent device prospect.\cite{Novoselov2004,wangqh-natnano-2012,Chhowalla-natchem-2013,zhangx-csr-2015} The in-plane anisotropic 2D materials, such as black phosphorus (BP)\cite{qiaojs-natcom-2014,liuh-acsn-2014,zhangyb-nnano-2014,wang2015highly}, SnSe\cite{nature-2014-SnSe,Science-2015-SnSe},rhenium disulfide (ReS$_2$)\cite{ReS2-2015-integrated} and Rhenium Diselenide (ReSe$_2$)\cite{wangh-2015}, have one more dimension to tune the properties than the isotropic 2D materials, like graphene and MoS$_2$\cite{Heinz-prl-2010}. The novel anisotropic properties of these systems make them be the promising materials for application in future electronics, optoelectronics and thermoelectrics.\cite{zhangyb-nnano-2014,Science-2015-SnSe,ReS2-2015-integrated} So far, anisotropic 2D materials can be divided into two categories by their structural properties. One category is the strongly buckled honeycomb sheet with 'troughs' running along the y-axis (b direction), and the typical materials are BP and SnSe.\cite{qiaojs-natcom-2014,ling2015low,nature-2014-SnSe} The other one lies disordered 1T-structures, which are unlike the common high symmetric 2H-structure (the common structure in transition metal dichalcogenides (TMDs), such as MoS$_2$, WSe$_2$\cite{WSe2-nn-2013-optical}). The classic 1T materials are ReS$_2$\cite{ReS2-2014-monolayer}, ReSe$_2$\cite{wangh-2015} and WTe$_2$\cite{WTe2-2014-large,prl-2014-electronic}.

The vdW interaction plays an important role in the physical properties of layered materials.\cite{Heinz-prl-2010,tanph-natm-2012,qiaoxf-apl-2015,lu2015rapid} For example, MoS$_2$ is tuned from direct to indirect bandgap semiconductor from monolayer to bilayer by the interlayer coupling.\cite{Heinz-prl-2010} Moreover, BP is a direct semiconductor with the bang gap varying from 0.3 of the bulk form to 1.5$\sim$2.0 eV of monolayer.\cite{BP-CSR-2015} The natural 2D materials usually stacking regularly, and the stacking order significantly affects the properties of 2D materials.\cite{wu-2014-resonant,lu2015rapid,puretzky2015low} For instance, the twisted bilayer graphene (BLG) emerges a pairs of singularities in the density of states in comparison with the Bernal-stacked BLG.\cite{dos-2007-graphene,wu-2014-resonant,wujb-acsn-2015} Understanding the interlayer coupling and layer stacking are of essential importance before stacking them to build 2D heterostructures for both device application and basic physical research.\cite{bonaccorso-2013-multiwall}

Anisotropic multilayer ReS$_2$ was previously found to be randomly stacked and lack of interlayer coupling,\cite{ReS2-2014-monolayer,ReS2-2015-integrated,PhysRevB.92.054110} as a consequence, bulk ReS$_2$ exhibits monolayer behavior due to electronic and vibrational decoupling in bulk ReS$_2$.\cite{ReS2-2014-monolayer} Here, two stable stacking orders in multilayer ReS$_2$, $aa$ and $\bar{a}b$ stackings, have been identified by Raman and photoluminescence (PL) spectroscopies, which are further confirmed by the density functional theory (DFT) calculations. Two shear (C) modes are observed in the $aa$-stacked bilayer (2L) ReS$_2$, indicating an anisotropic structure. However, the $\bar{a}b$-stacked 2L-ReS$_2$ exhibits an isotropic feature because only one C mode is observed in the $\bar{a}b$-stacked 2L-ReS$_2$. The observation of ultralow-frequency interlayer vibration modes suggests that multilayer ReS$_2$ with both the two stacking orders shows unexpected significant interlayer couplings, the strengths of which are 55-90\% to those of multilayer MoS$_2$. Moreover, The two stacking ways in multilayer ReS$_2$ can be clearly identified by their high-frequency Raman modes and PL spectra. The discovery of strong interlayer coupling and polytypism in multi-layer ReS$_2$ open the possibility to further understand the layer stacking behaviors in other 2D anisotropic materials.
\begin{figure} \centerline{\includegraphics[width=160mm,clip]{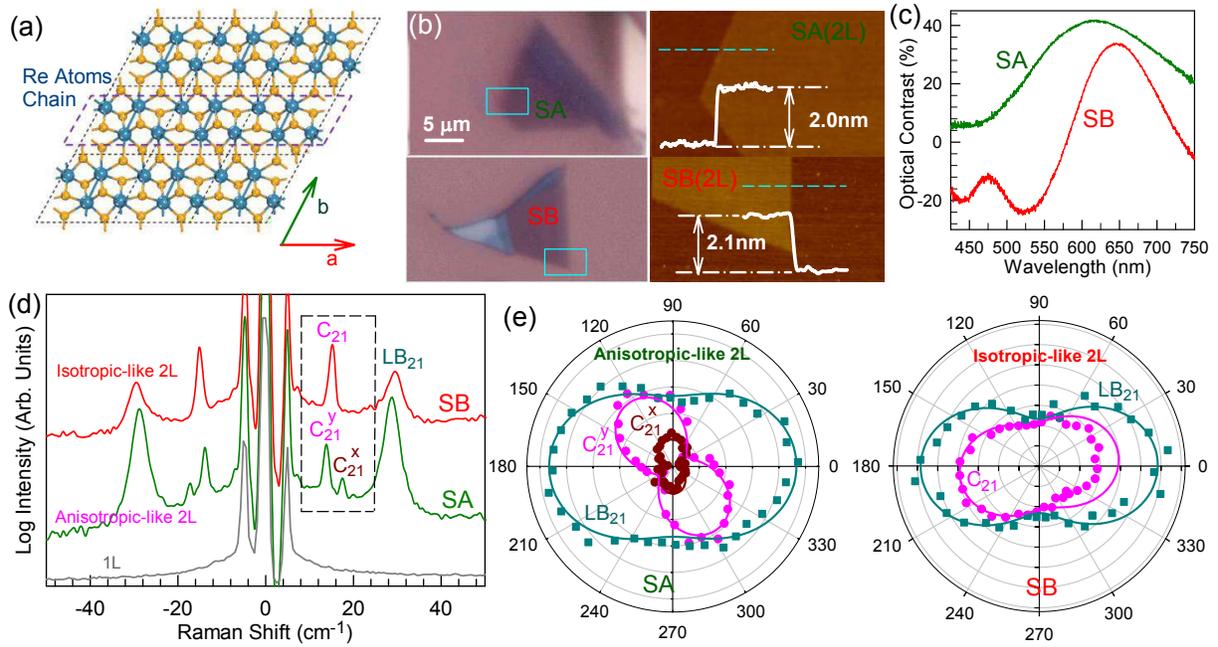}} \caption{({\bf a}) Crystal structure of 1L ReS$_2$ with 3$\times$3 unit cells. The Re atom chain along \textbf{\emph{a}} direction is marked by the dash frame.({\bf b}) Optical image of two ReS$_2$ flakes (SA and SB) on a 90 nm SiO$_2$/Si substrate and the corresponding AFM images within the square frames of the two flakes. The height profiles indicate that both SA and SB flakes are 2L-ReS$_2$. ({\bf c}) Optical contrast of the flakes SA and SB. ({\bf d}) Stokes/anti-Stokes Raman spectra of the flakes SA and SB in the ULF region. ({\bf e}) The intensities of the C and LB modes for the flakes SA and SB as a function of laser polarization angle. Circles and Squares are experimental data and solid curves are the trends based on the symmetry analysis.} \label{Fig1} \end{figure}

Figure 1(a) shows the crystal structure of the monolayer ReS$_2$, whose unit cell contains four formula units consisting of two categories of rhenium (Re) atoms together with four categories of sulfur (S) atoms. Each Re in monolayer ReS$_2$ has six neighboring S sites, and the Re atoms are sandwiched by the S atoms at the both sides. Unlike common TMDs such as MoS$_2$ and WSe$_2$ which crystallize in the hexagonal (H) phases, ReS$_2$ crystals exhibit a distorted CdCl$_2$-type lattice structure\cite{lamfers1996crystal}. Owing to the Peierls distortion\cite{kertesz1984octahedral}, four adjacent Re atoms are bonded together to form a zigzag chain \cite{fang1997electronic}, oriented along the direction of lattice vector \textbf{\emph{a}} (see Fig. 1(a)). Figure 1(b) shows two 2L ReS$_2$ flakes, SA and SB. The atomic force microscopy (AFM) measurement in Fig. 1(b) reveals the thicknesses of the flakes SA and SB are 2.0 and 2.1 nm, respectively, which are about 0.7 nm larger than the expected thickness (1.4 nm) of 2L ReS$_2$. This discrepancy is most likely ascribed to the  instrumental offset for the absolute values. Indeed, the thickness ($\sim$0.7 nm) of 1L ReS$_2$ was measured to be 1.4 nm as demonstrated in Fig. S1 in the SI. Although the SA and SB flakes are identified as 2L ReS$_2$, they exhibit different optical contrast, suggesting that 2L ReS$_2$ can crystallize in different crystal structures, in contrast to the previous reports\cite{ReS2-2014-monolayer,ReS2-2015-integrated,PhysRevB.92.054110}.

1L and bulk ReS$_2$ belong to the $C_i$ space group. The unit cell of 1L ReS$_2$ is comprised of 12 atoms with 36 normal vibrational modes. In principle, for N-layer (NL) ReS$_2$ with a unit cell of 12N atoms, 36N modes are expected. The $\Gamma$ phonons of 1L ReS$_2$ can be expressed by the irreducible representations of $C_i$ as follows: $\Gamma$ = 18(A$''$+ A$'$), where three A$''$ are acoustic modes, the other A$''$ are infrared active, all the 18 A$'$ modes are Raman active. The Raman spectrum of 1L ReS$_2$ has been studied by several groups\cite{ReS2-2014-monolayer,PhysRevB.92.054110,Nagler-2015-C} and 18 A$'$ modes are already observed. As depicted in the Fig. S2 in the Supplementary Information, more than 18 Raman modes are observed in the Raman spectra of the flakes SA and SB in the range of 100-450 cm$^{-1}$, which are similar to each other. This indicates that the SA and SB flakes have the similar in-plane atomic structures.

Interlayer vibration modes were used to characterize the interlayer coupling\cite{tanph-natm-2012,zhangx-prb-2013,zhangx-csr-2015,wangh-2015} and the stacking orders\cite{wu-2014-resonant,wujb-acsn-2015,lu2015rapid,puretzky2015low,zhangxin-Carbon-2016} of multilayer flakes of 2D materials. In particular, the shear (C) modes are very sensitive with the stacking condition.\cite{wu-2014-resonant} For example, in the AB-stacked bilayer graphene (BLG)\cite{tanph-natm-2012}, the C mode is at 32 cm$^{-1}$, but it cannot be observed in the twisted bilayer graphene (tBLG)\cite{wu-2014-resonant}, due to the periodicity mismatch between two twisted layers. However, the twisting at the interface of twisted multilayer graphenes (MLG) does not substantially modify the frequency of the layer breathing (LB) modes because the interlayer distance is nearly unchanged after twisting.\cite{wujb-acsn-2015} On the other hand, in the twisted 2L-MoS$_2$, the LB modes slightly soften from that of AB-stacked 2L-MoS$_2$,\cite{tBL-MoS2} due to the twisting induced increase of the interlayer distance\cite{tBL-MoS2,jiang2014valley}. For in-plane isotropic 2D materials, such as MLG\cite{tanph-natm-2012} and multilayer MoS$_2$\cite{zhangx-prb-2013,zhao2013interlayer}, the C mode is doubly degenerate. However, in an anisotropic 2D material, such as BP\cite{kaneta1986lattice,ling2015low}, the C modes are non-degenerated and each pair of the C modes should be denoted as $C^x$ and $C^y$. For an NL 2D flake, there are N-1 pairs of C modes and N-1 LB modes. We denote the C and LB modes with the highest frequency, respectively, as C$^x_{N1}$, C$^y_{N1}$ and LB$_{N1}$, and those with the second highest frequency as C$^x_{N2}$, C$^y_{N2}$ and LB$_{N2}$.

Ultralow-frequency (ULF) Raman spectroscopy\cite{tanph-natm-2012} is used to further identify the interlayer coupling and the stacking orders ReS$_2$ flakes SA and SB, as shown in Fig. 1(d). Three peaks are located at 13.8, 17.5 and 28.7 cm$^{-1}$ in the flake SA. However, only two peaks are observed at 15.1 and 29.3 cm $^{-1}$ in the flake SB. There is no ULF mode in the 1L ReS$_2$ flake whose optical image is shown in Fig. S1 of SI, further confirming that the flake is monolayer due to the absence of the C and LB modes\cite{tanph-natm-2012}. Because the LB coupling is usually stronger than the shear coupling in layered materials, we identify the mode with the highest frequency in flakes SA and SB as the LB$_{21}$ mode. We further identify the modes at 13.8 and 17.5 cm$^{-1}$ in flake SA as C$^y_{21}$ and C$^x_{21}$, respectively, and the mode at 15.1 cm $^{-1}$ in flake SB as C$_{21}$. Two C modes observed in the SA flake indicate its anisotropic-like feature, similar to the BP case.\cite{kaneta1986lattice,ling2015low} However, only one C mode is observed in the SB flake, suggesting its isotropic-like feature, similar to the case of multilayer MoS$_2$ and MLG\cite{tanph-natm-2012,zhangx-prb-2013}. All of the interlayer C and LB modes shows a polarization dependence on the mode intensity, as depicted in Fig. 1(e). This polarization dependence behavior directly results from its low crystal symmetry ($C_i$) although the interlayer vibration is isotropic-like in the SB flake of 2L-ReS$_2$.

\begin{figure} \centerline{\includegraphics[width=120mm,clip]{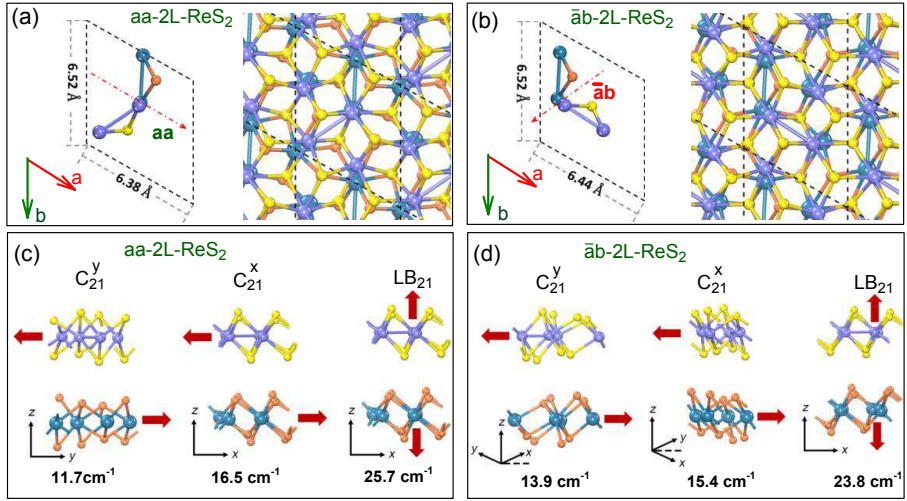}} \caption{({\bf a,b}) Stacking schematic diagram and top view of crystal structure of the aa-stacked ({\bf a}) and $\bar{a}b$-stacked ({\bf b}) 2L-ReS$_2$. The rotation axes are labeled by red dot-dashed lines. ({\bf c,d}) Atomic displacement schematic of three ULF interlayer modes in aa-stacked ({\bf c}) and $\bar{a}b$-stacked ({\bf d}) 2L-ReS$_2$ obtained from the DFPT calculation. The corresponding frequencies are marked.} \label{Fig2}
\end{figure}

Both the optical contrast and measured frequencies of interlayer vibration modes indicate that the SA and SB flakes are 2L-ReS$_2$ with different crystal structural symmetries or layer stacking orders. As shown in Fig. 1(a), a formed Re-Re bond is almost oriented parallel to the \textbf{\emph{b}} direction, and the top S atom bonding with these two Re atoms are in the forward side of the \textbf{\emph{a}} direction. This Re-S-Re triangle can be adopted to illustrate the crystal structural symmetry of NL ReS$_2$. There are many stacking orders of 2L-ReS$_2$ built from two 1L-ReS$_2$ by the in-plane relative rotation and translation. In each case, the top layer can almost superpose with the bottom layer after the top layer is rotated 180 $^{\circ}$ along a special in-plane rotation axis and then shifted along in-plane direction. Since the crystal structure of 1L ReS$_2$ is reduced from the hexagonal crystal system, there are 6 in-plane rotation axes. Because of the in-plane anisotropy, the 6 in-plane rotation axes increase to 12 axes after the axis direction is considered, where the orientation along the opposite direction of the included angle of the two Re-Re bonds can be set as the positive direction, see Fig. 2(a)(b). Fig. 2(a)(b) demonstrate the possible configuration with rotation axes of \textbf{\emph{aa}} and \textbf{$\bar{a}b$}, and the corresponding 2L ReS$_2$ are denoted as aa-stacked 2L-ReS$_2$ (aa-2L-ReS$_2$) and $\bar{a}b$-stacked 2L-ReS$_2$ ($\bar{a}b$-2L-ReS$_2$), respectively. The axis \textbf{\emph{aa}} is in the same direction with the vector \textbf{\emph{a}}, while the axis \textbf{$\bar{a}b$} is in the same direction with the vector -\textbf{\emph{a}}+\textbf{\emph{b}}. In general, in this quasi-hexagonal crystal system, there are 12 possible stacking ways, as shown in Fig. S3 in the SI. The 12 possible stacking ways for 2L-ReS$_2$ are relaxed by density functional theory (DFT) (see the method for detail), in which relative small translation between the top and bottom ReS$_2$ layers for each stacking way is considered. The free relaxation of crystal structures show that the top and bottom ReS$_2$ layers in 2L-ReS$_2$ tends to not shift for each stacking way. The aa-2L-ReS$_2$ is the most stable configuration with the lowest total energy (referenced to 0 eV), and the $\bar{a}b$-2L-ReS$_2$ is the second most stable configuration with the second lowest energy (0.09 eV). After the free relaxation by DFT, the lattice constants of aa-2L-ReS$_2$ are 6.38 (|\textbf{\emph{a}}|) and 6.52 (|\textbf{\emph{b}}|) \AA, and the lattice constants of $\bar{a}b$-2L-ReS$_2$ are 6.44 (|\textbf{\emph{a}}|) and 6.52 (|\textbf{\emph{b}}|) \AA.

In a bilayer 2D system, the frequency difference between $C^x$ and $C^y$ is dependent on the lattice constant difference between the direction \textbf{\emph{a}} and \textbf{\emph{b}}. For example, the |\textbf{\emph{a}}| and |\textbf{\emph{b}}| are 3.32 and 4.58 \AA in BP,\cite{qiaojs-natcom-2014} which leads to that the frequencies of C$^x$ and C$^y$ in 2L-BP are 13.7 and 36.5 cm$^{-1}$.\cite{kaneta1986lattice,ling2015low} The difference between |\textbf{\emph{a}}| and |\textbf{\emph{b}}| of aa-2L-ReS$_2$ is bigger than that of $\bar{a}b$-2L-ReS$_2$, which indicates that the frequency difference between the C$^x$ and C$^y$ modes of aa-2L-ReS$_2$ is bigger than that of the $\bar{a}b$-2L-ReS$_2$. The frequency difference of the two C modes in the flake SA is 3.7 cm$^{-1}$, and that in the flake SB is expected to be small so that only one C mode observed in the flake SB. Thus, we identify the flake SA as aa-2L-ReS$_2$, and the flake SB as $\bar{a}b$-2L-ReS$_2$.

As shown in Fig. 2(c), the calculated frequency of the C$^y_{21}$, C$^x_{21}$ and LB$_{21}$ modes, Pos(C$^y_{21}$), Pos(C$^x_{21}$) and Pos(LB$_{21}$), for the aa-2L-ReS$_2$ are 11.7, 16.5 and 25.7 cm$^{-1}$, respectively, which are closed to the experimental results of 13.8, 17.5 and 28.7 cm$^{-1}$. The theoretical Pos(C$^y_{21}$), Pos(C$^x_{21}$) and Pos(LB$_{21}$) for $\bar{a}b$-2L-ReS$_2$ are (13.9, 15.4) and 23.8 cm$^{-1}$, respectively, as shown in Fig. 2(d), also consistent with the experimental values of 15.1 and 29.3 cm$^{-1}$. Moreover, 4.8 cm$^{-1}$ of Pos(C$^x_{21}$)-Pos(C$^y_{21}$) in the aa-2L-ReS$_2$ agrees with the experimental value of 3.7 cm$^{-1}$. 1.5 cm$^{-1}$ of Pos(C$^x_{21}$)-Pos(C$^y_{21}$) in the $\bar{a}b$-2L-ReS$_2$ is comparable with the full width at half maximum (FWHM) of 1.3-1.9 cm$^{-1}$ for the C mode, making it hard to be resolved in the experiment. These results confirm the identification of the stacking orders for the flakes SA and SB. Because aa-2L-ReS$_2$ is expected to be the most stable configuration, the aa-2L-ReS$_2$ is more frequently exfoliated in the experiment. Indeed, the 80\% of the exfoliated 2L-ReS$_2$ exhibit the two C modes in the ULF Raman spectra.

\begin{figure} \centerline{\includegraphics[width=160mm,clip]{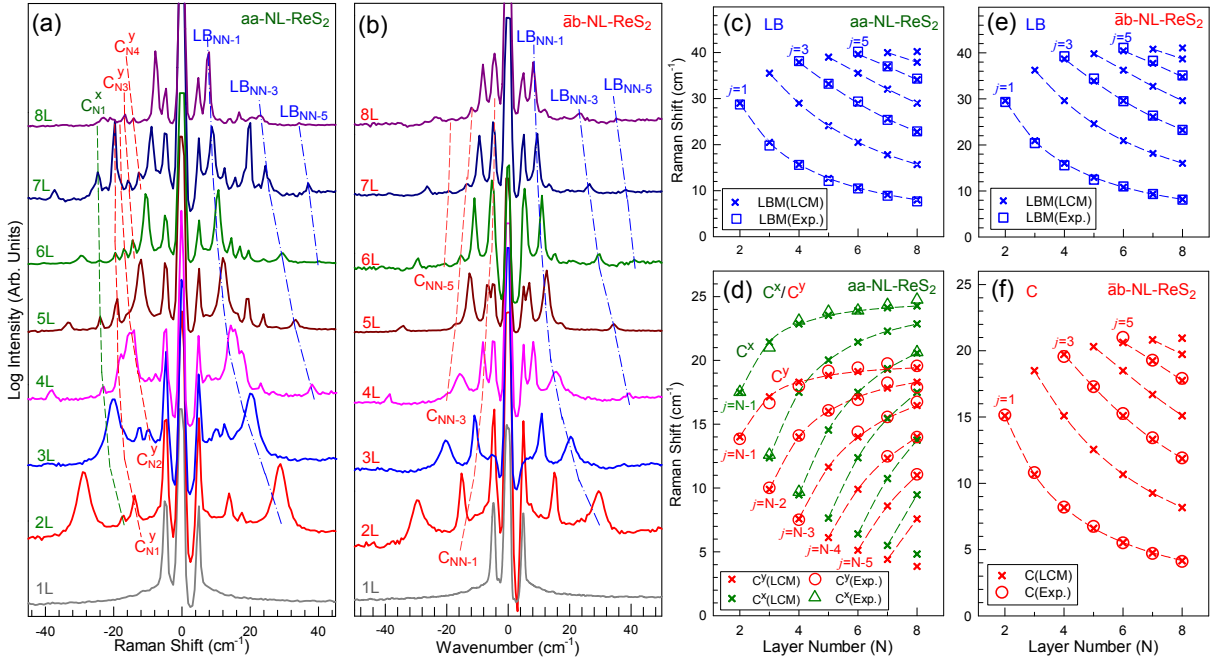}} \caption{({\bf a,b}) Stokes/anti-Stokes Raman spectra in the C and LB peak region for aa-stacked ({\bf a}) and $\bar{a}b$-stacked ({\bf b}) 2-8L ReS$_2$ along with that of 1L ReS$_2$. The dashed lines are guides to the eye. ({\bf c,d}) Positions of LB ({\bf c}) and C ({\bf d}) modes as a function of N for aa-stacked 2-8L ReS$_2$. The blue open squares, green open triangles and red open circles are the experiment data. The crosses are the theoretical data based on LCM. The index of each branch observed in the experiment is labeled. ({\bf e,f}) Positions of LB ({\bf e}) and C ({\bf f}) modes as a function of N for aa-stacked 2-8L ReS$_2$. The blue open squares and red open circles are the experiment data. The crosses are the theoretical data based on LCM. The index of each branch observed in the experiment is labeled.} \label{Fig3}
\end{figure}

ULF Raman spectroscopy is further applied for NL ReS$_2$ (N$>$2). We find that for each NL-ReS$_2$, the ULF Raman spectra can be divided into two categories, which exhibit different spectral features in mode number and peak position. After analyzing the peak position for each mode in these two categories, we find that the two categories of NL ReS$_2$ correspond to the aa and $\bar{a}b$ stacking configurations as addressed above in 2L-ReS$_2$, and are denoted as aa-NL-ReS$_2$ and $\bar{a}b$-NL-ReS$_2$ (N$>$1), respectively. The ULF Raman spectra of aa-NL-ReS$_2$ and $\bar{a}b$-NL-ReS$_2$ are depicted in Fig.3(a) and Fig.3(b), respectively, together with the data from 1L ReS$_2$ presented for comparison. The linear chain mode (LCM) can be used to describe the N-dependent interlayer mode frequency where only the nearest-neighbor interlayer interaction is considered.\cite{tanph-natm-2012} The frequency of N-1 LB modes of in NL ReS$_2$ can be given as follows\cite{zhangx-prb-2013}:

\begin{equation} Pos(LB_{N,N-j})=\sqrt{2}Pos(LB_{21})\sin(j\pi/2N), \end{equation}

\noindent where $j$ is an integer, $j$ = N-1, N-2,..., 2,1. The above equation can also be applied to the C modes in the aa-NL-ReS$_2$ (N$>$1) by replacing $Pos(LB_{N,N-j})$ with $Pos(C^y_{N,N-j})$ and $Pos(C^x_{N,N-j})$, and the C modes in the $\bar{a}b$-NL-ReS$_2$ by replacing $Pos(LB_{N,N-j})$ with $Pos(C_{N,N-j})$. Once the frequencies of interlayer modes of 2L ReS$_2$ are determined, those in NL ReS$_2$ can be predicted. As demonstrated in the Fig. 3(a) for the aa-NL-ReS$_2$, the LB$_{NN-1}$, LB$_{NN-3}$ and LB$_{NN-5}$ branches are marked by the blue dash-dotted lines, and the C$^x_{N1}$, C$^y_{N1}$, C$^y_{N2}$, C$^y_{N3}$, C$^y_{N4}$ branches are marked by the green or red dashed lines, respectively. The prominent interlayer mode branches (C$_{NN-1}$, C$_{NN-3}$, C$_{NN-5}$, LB$_{NN-1}$, LB$_{NN-3}$, LB$_{NN-5}$) of $\bar{a}b$-NL-ReS$_2$ are also demonstrated in the Fig. 3(b). The spectral feature is quite complex and peak fitting is necessary for certain aa-NL-ReS$_2$ to identify some interlayer modes, as depicted in Fig. S4 in the SI for aa-3L-ReS$_2$ and aa-4L-ReS$_2$. All the experimental Pos(C) and Pos(LB) of aa-NL-ReS$_2$ are summarized in Fig. 3(c) and 3(d), and those for $\bar{a}b$-NL-ReS$_2$ in Fig. 3(e) and 3(f), respectively. All of them are in good agreement with the theoretical results based on LCM.

As shown in Fig. 3(c)(e), the observed LB branches for aa and $\bar{a}b$ stacking configurations in NL ReS$_2$ are $j$=N-1, N-3 and N-5. The interlayer mode in each branch of $j$=N-1, N-3 or N-5 increases in frequency with increasing N, which is also true for the C branches in $\bar{a}b$-NL-ReS$_2$. However, the observed C branches in aa-NL-ReS$_2$ are $j$=1 for C$^x$ and $j$=1, 2, 3 and 4 for C$^y$. It may result from the different symmetries and electron-phonon couplings between the aa and $\bar{a}b$ stacking orders.\cite{wu-2014-resonant} Fig. 3(c-f) has shown that N-dependent Pos(C) and Pos(LB) in aa- and $\bar{a}b$-NL-ReS$_2$ can be well explained by the LCM. The interlayer coupling can be deduced from the mode frequency of the C and LB modes of 2L-ReS$_2$ by a formula\cite{zhangx-prb-2013} of $\alpha_C=(2\pi^{2}c^{2})\mu Pos(C_{21})$ and $\alpha_{LB}=(2\pi^{2}c^{2})\mu Pos(LB_{21})$, where $\alpha_C$ and $\alpha_{LB}$ is the interlayer shear and breathing force constant per unit area, $\mu$ atomic mass per unit area, and $Pos{C_{21}}$ and $Pos{LB_{21}}$ the frequency of the C and LB modes of 2L-ReS$_2$, respectively. C$^x_{21}$, C$^y_{21}$ and LB$_{21}$ in aa-2L-ReS$_2$ are observed at 17.5, 13.8 and 28.7 cm$^{-1}$, respectively. These frequencies are slightly smaller than those\cite{zhangx-prb-2013} of the C (22.6 cm$^{-1}$) and LB (40.1 cm$^{-1}$) modes of 2L-MoS$_2$. Because $\mu$(ReS$_2$) is about 1.5 times to$\mu$(MoS$_2$), $\alpha_C$ of anisotropic-like aa-2L-ReS$_2$ along X and Y axes are about 90\% and 55\% to those of 2L-MoS$_2$, respectively, significantly anisotropic in the basal plane. $\alpha_{LB}$ in 2L-ReS$_2$ is about 76\% to that in 2L-MoS$_2$. For the isotropic-like $\bar{a}b$-2L-ReS$_2$, $\alpha_C$ is about 67\% of that in 2L-MoS$_2$. Our results show compelling evidence of significant interlayer coupling in aa- and $\bar{a}b$-NL-ReS$_2$, which are comparable with those in multilayer MoS$_2$.

\begin{figure} \centerline{\includegraphics[width=160mm,clip]{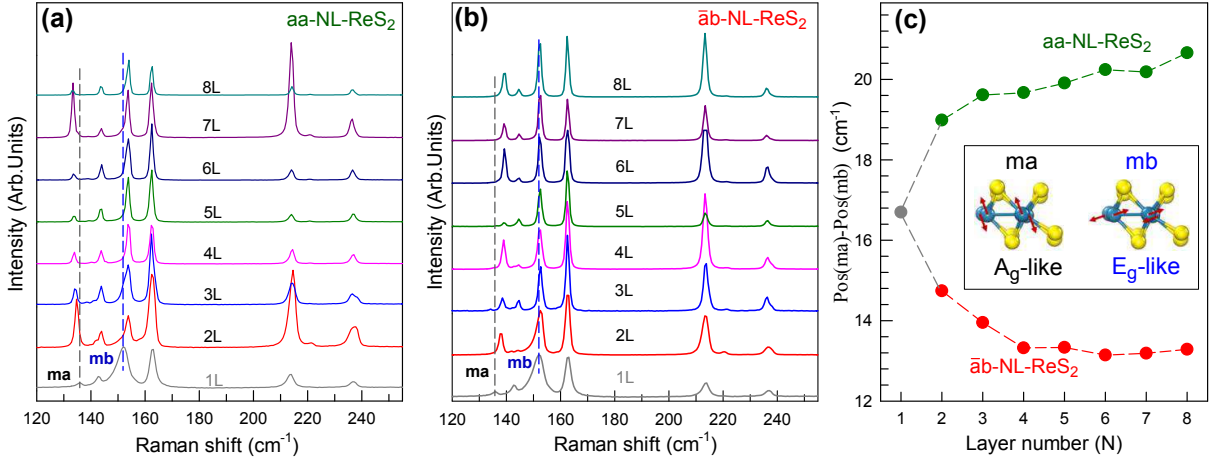}} \caption{({\bf a,b}) Raman spectra of aa-stacked ({\bf a}) and $\bar{a}b$-stacked ({\bf b}) 2-8L ReS$_2$ in the range of 120-250 cm$^{-1}$, together with that of 1L ReS$_2$ for comparison. The dashed lines are guides to the eye. ({\bf c}) Pos(mb)-Pos(ma) as a function of N for aa-stacked (green circles) and $\bar{a}b$-stacked (red circles) 2-8L ReS$_2$. The inset shows the atoms displacement of the modes ma and mb, which are indicated in ({\bf a}) and ({\bf b}).}\label{Fig4}
\end{figure}

The N-dependent high-frequency Raman modes also reflect the interlayer interaction in 2D materials.\cite{zhangx-csr-2015} The optical Raman spectra of aa- and $\bar{a}b$-NL-ReS$_2$ up to 8L are shown in Figs. 4(a) and 4(b) in the high frequency region from 120-250 cm$^{-1}$, respectively, together with that of 1L ReS$_2$. With increasing N from 1L, NL ReS$_2$ exhibit a similar spectral feature, including mode number and mode position. Because each mode has specific intensity dependence on the laser polarization direction, the relative intensity between different modes is usually sample dependent. The peaks at $\sim$ 161, 213, 235 cm$^{-1}$ do not obviously change with N for both aa- and $\bar{a}b$-NL-ReS$_2$. As shown by the dashed lines in Fig. 4, the ma and mb peaks exhibit different N-dependent peak position from 1L ReS$_2$ to aa-8L-ReS$_2$ or $\bar{a}b$-8L-ReS$_2$. With increasing N, the ma mode of 1L ReS$_2$ softens in aa-stacked NL-ReS$_2$ while it stiffens for $\bar{a}b$-stacked ones. However, the mb mode stiffens in aa-stacked multilayers but almost keeps constant for all $\bar{a}b$-stacked samples. Consequently, Pos(mb)-Pos(ma) increases from 16.4 cm$^{-1}$ of 1L ReS$_2$ to 20.6 cm$^{-1}$ of aa-8L-ReS$_2$, while it decreases from 16.4 cm$^{-1}$ of 1L ReS$_2$ to 13.1 cm$^{-1}$ of $\bar{a}b$-8L-ReS$_2$, as demonstrated in Fig. 4(c). Therefore, Pos(mb)-Pos(ma) can be used to identify N and the stacking way, similar to the case of few-layer MoS$_2$.\cite{zhangx-csr-2015}

The inset in Fig. 4(c) shows the atomic displacement of the ma and mb modes. For a traditional TMD, such as MoS$_2$, we denote the mode with large weight of out-of-plane vibrations as the $A_g$-like mode, and the one with large weight of in-plane vibrations as the $E_g$-like mode.\cite{molina2011phonons} Thus, the $ma$ and $mb$ modes are the $A_g$-like and $E_g$-like modes, respectively. In general, the intra-layer mode can be stiffened by the interlayer vdW coupling and be softened by the long-range Coulomb screen.\cite{molina2011phonons} The overall strength between the two interactions determines the mode to be stiffened or softened with increasing N. Both the interlayer coupling and long-range Coulomb screen are sensitive with the stacking ways. The calculated frequencies of the $ma$ and $mb$ modes of 1L ReS$_2$, aa-2L-ReS$_2$ and $\bar{a}b$-2L-ReS$_2$ by DFPT are summarized in Table S1 in the SI. The calculated Pos(mb)-Pos(ma) increases from 17.9 cm$^{-1}$ of 1L ReS$_2$ to 18.2 cm$^{-1}$ of aa-2L-ReS$_2$, while it decreases to 16.9 cm$^{-1}$ for $\bar{a}b$-2L-ReS$_2$, consistent with the experimental trend. Therefore, the calculated and experimental Pos(mb)-Pos(ma) values in 1L- and 2L-ReS$_2$ further confirm that the significant interlayer coupling really exists in NL ReS$_2$ and is different between the two stacking orders.

\begin{figure} \centerline{\includegraphics[width=120mm,clip]{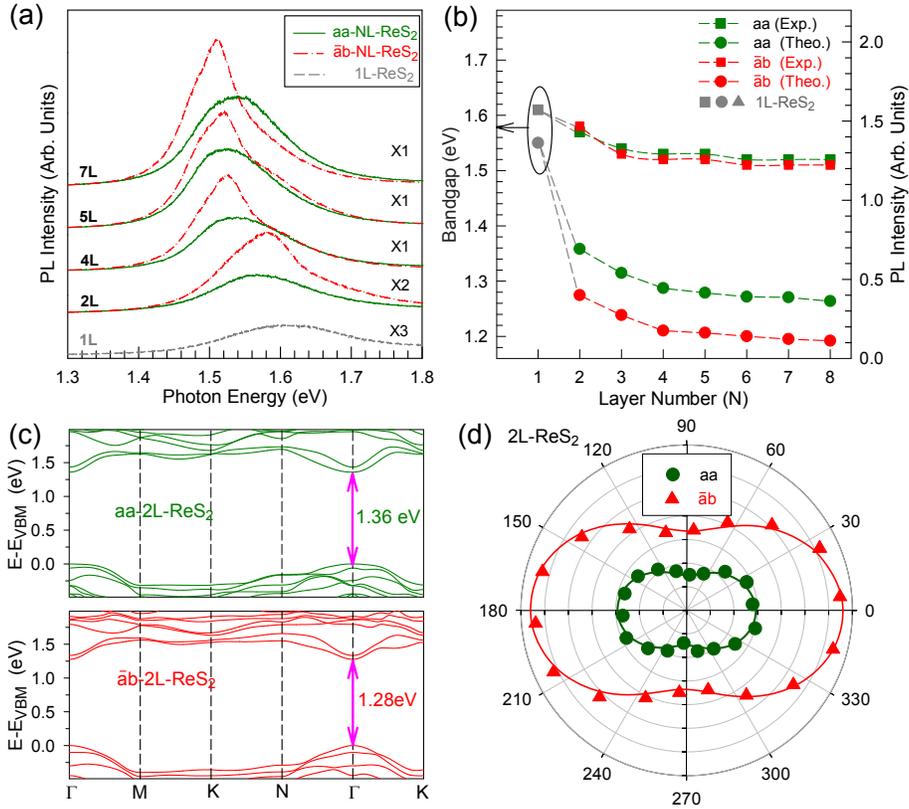}} \caption{({\bf a}) PL spectra of 2L, 4L, 5L and 7L ReS$_2$ with aa (green curve) and $\bar{a}b$ (red curve) stacking ways. ({\bf b}) Evolution of direct bandgap of aa-stacked (green) and $\bar{a}b$-stacked (red) NL-ReS$_2$ as a function of N. The bandgap is extracted from PL spectra (square) and also revealed by DFT calculation (circle). ({\bf c}) Band structures of aa-stacked (top panel, green solid curves) and $\bar{a}b$-stacked (bottom panel, red solid curves) 2L-ReS$_2$, respectively, where the corresponding bandgap is marked. ({\bf d}) The polarization dependence of the PL intensity of aa-stacked (green) and $\bar{a}b$-stacked (red) 2L-ReS$_2$.} \label{Fig5}
\end{figure}

The bandgaps of aa- and $\bar{a}b$-NL-ReS$_2$ are measured by PL spectra, as shown in Fig. 5(a). Figure 5(a) also shows the PL spectrum of 1L ReS$_2$ in the grey curve, revealing its bandgap of 1.61 eV. In general, for a given N, PL spectra of aa-NL-ReS$_2$ (green solid curves in Fig. 5(a)) exhibit weaker intensity and higher peak energy in comparison with $\bar{a}b$-NL-ReS$_2$. The PL profile of aa-NL-ReS$_2$ is similar to that of 1L ReS$_2$ and with a full width at half maximum (FWHM) of about 160 meV. However, PL profile of $\bar{a}b$-NL-ReS$_2$ is narrower than that of aa-NL-ReS$_2$, and there exists a sharp peak at its center. These different PL profiles can be adopted as a gauge to distinguish the $\bar{a}b$ stacked NL ReS$_2$ from the aa-stacked ones. The bandgap of NL ReS$_2$ decreases from 1.57 eV of 2L to 1.52 eV of 8L for aa-NL-ReS$_2$ while it decrease from 1.58 eV of 2L to 1.51 eV of 8L for $\bar{a}b$-NL-ReS$_2$, as summarized in Fig. 5(b). The calculated band structures show that 1L ReS$_2$ and both aa- and $\bar{a}b$-NL-ReS$_2$ are direct bandgap semiconductors, corresponding to the transition at the $\Gamma$ point of the Brillouin zone, as shown in Fig. 5(c). The theoretical bandgap of 1L ReS$_2$ is 1.55 eV, which drops to 1.36 and 1.27 eV for the aa- and $\bar{a}b$-stacked bilayers, and further to 1.26 eV of aa-8L-ReS$_2$ and 1.19 eV of $\bar{a}b$-8L-ReS$_2$. The theoretical and experimental trends of bandgap on N are consistent with the only exception for 2L-ReS$_2$ that experimental bandgap of aa-2L-ReS$_2$ is smaller than that of $\bar{a}b$-2L-ReS$_2$, but the calculation predicts the opposite. Standard DFT is widely known to underestimate bandgap of semiconductors and does not consider exciton binding energy that is of importance in 2D layered structures. In terms of TMDs, standard DFT predicts reasonable bandgaps because of an error cancelation by the underestimation and lack of exciton binding energy. In thinner flakes, e.g. 2L-ReS$_2$, the role of exciton binding energy is more pronounced than those of thicker flakes. We thus argue that this experiment-theory discrepancy in 2L-ReS$_2$ is, most likely, a consequence of different exciton binding energies of aa- and $\bar{a}b$-2L-ReS$_2$. Figures 5(d) shows the polarized PL intensities of both aa- and $\bar{a}b$-2L-ReS$_2$, which exhibit comparable polarization behaviors because the similar in-plane crystal structures. The N-dependent PL peaks also indicate that the interlayer coupling in multi-layer ReS$_2$ is significant for both stacking orders.

In conclusion, two stable stacking orders (aa and $\bar{a}b$) in the 2L ReS$_2$ has been identified by ultralow-frequency Raman spectroscopy and the first-principles calculation, which indicates the significant interlayer coupling in 2L ReS$_2$, in different from the previously believed non-interaction. According to N-dependent frequency evolution of the C and LB modes, the aa and $\bar{a}b$ stacking orders also exist in multi-layer ReS$_2$ and the corresponding N-dependent frequency of C and LB modes can be well predicted by the linear chain model. The N-dependent high-frequency Raman modes and PL spectra further confirm the existence of the aa and $\bar{a}b$ stacking orders in multi-layer ReS$_2$. The isotropic interlayer shear coupling with the appearance of one C mode, Pos(mb)-Pos(ma) and peak profile can be used to distinguish $\bar{a}b$-stacked NL-ReS$_2$ from aa-stacked ones. This study pave the way to explore interlayer coupling, layer stacking, electronic and optical properties of other 2D anisotropic materials.

\section{Methods}
\subsection{Sample}
The ReS$_2$ flakes were prepared on Si/SiO$_2$ substrates from a bulk crystal (2D semiconductors Inc) by the standard micromechanical exfoliation method.\cite{Novoselov2004}

\subsection{Optical contrast}
Micro-Raman confocal system and a $\times$50 objective with a numerical aperture of 0.45 were used in the optical contrast measurement. A tungsten-halogen lamp was used as a light source. The optical contrast is defined as 1-R$_F$($\lambda$)/R$_{Sub}$($\lambda$), where R$_{Sub}$($\lambda$) and R$_F$($\lambda$) are the reflected light intensities from the SiO$_2$/Si bare substrate and the ReS$_2$ flake deposited on SiO$_2$/Si substrate, respectively, dependent on the wavelength ($\lambda$) of the light source.

\subsection{Raman and photoluminescence spectroscopies}
Raman and PL spectra are measured at room temperature using a Jobin-Yvon HR800 micro-Raman system equipped with a liquid-nitrogen-cooled charge couple detector (CCD), a $\times$100 objective lens (numerical aperture=0.90) and several gratings. The excitation wavelengths are 502 nm from an Ar$^+$ laser and 633 nm from a He-Ne laser. The 1800 lines/mm grating is used in the Raman measurement, which enables one to have each CCD pixel to cover 0.60 cm$^{-1}$ at 502 nm. A typical laser power of 0.3 mW is used to avoid sample heating. The laser plasma lines are removed using a BragGrate bandpass filter. Measurements down to 5 cm$^{-1}$ are enabled by three BragGrate notch filters with optical density 4 and with a FWHM of 5 cm$^{-1}$.

\subsection{Density functional theory calculation} Density functional theory calculations were performed using the generalized gradient approximation for the exchange-correlation potential, the projector augmented wave method\cite{Blochl-prb-1994,Kresse-prb-1999} and a plane-wave basis set as implemented in the Vienna ab-initio simulation package (VASP)\cite{Kresse-prb-1996}. Phonon-related properties were calculated with VASP based on density functional perturbation theory (DFPT)\cite{Baroni-Rmp-2001}. The energy cut-off for the plane-wave basis was set to 700 eV for all calculations. A 2$\times$2 supercell containing two ReS$_2$ layers and a vacuum layer of 16 \r{A} was adopt to consider the distorted lattice with a k-mesh of 7$\times$7$\times$1 to sample the first Brillouin zone of few-layered ReS$_2$. Van der Waals interactions were considered at the vdW-DF\cite{Dion-prl-2004,LeeK-prb-2010} level with the optB86b exchange functional (optB86b-vdW)\cite{Klimes-jp-2010,Klimes-prb-2011} for geometry optimization, which was found to be more accurate in describing structural properties of layered materials\cite{huzx-sp-2014,hongjh-natcommu-2015,wujb-acsn-2015,huzx-ns-2015}. The shape (in-plane lattice parameters) of each supercell was fully optimized and all atoms in the supercell were allowed to relax until the residual force per atom was less than 0.001 eV$\cdot$\AA$^{-1}$. In electronic bandstructure calculations, the lattice constants were kept fixed to the 2L values for multilayers, and the residual forces of all atoms in each supercell were fully relaxed to less than 0.01 eV$\cdot$\AA$^{-1}$.

\emph{Note: While finalizing this manuscript, we became aware of two preprints reporting the low-frequency Raman spectra of ReS$_2$.\cite{he2015coupling,lorchat2015splitting}}

\begin{acknowledgement} This work was supported by the National Natural Science Foundation of China, Grant Nos. 11225421, 11434010, and 11474277.
\end{acknowledgement}
\bibliography{ReS2-ArXiv}

\providecommand*\mcitethebibliography{\thebibliography}
\csname @ifundefined\endcsname{endmcitethebibliography}
  {\let\endmcitethebibliography\endthebibliography}{}
\begin{mcitethebibliography}{53}
\providecommand*\natexlab[1]{#1}
\providecommand*\mciteSetBstSublistMode[1]{}
\providecommand*\mciteSetBstMaxWidthForm[2]{}
\providecommand*\mciteBstWouldAddEndPuncttrue
  {\def\EndOfBibitem{\unskip.}}
\providecommand*\mciteBstWouldAddEndPunctfalse
  {\let\EndOfBibitem\relax}
\providecommand*\mciteSetBstMidEndSepPunct[3]{}
\providecommand*\mciteSetBstSublistLabelBeginEnd[3]{}
\providecommand*\EndOfBibitem{}
\mciteSetBstSublistMode{f}
\mciteSetBstMaxWidthForm{subitem}{(\alph{mcitesubitemcount})}
\mciteSetBstSublistLabelBeginEnd
  {\mcitemaxwidthsubitemform\space}
  {\relax}
  {\relax}

\bibitem[Novoselov et~al.(2004)Novoselov, Geim, Morozov, Jiang, Zhang, Dubonos,
  Grigorieva, and Firsov]{Novoselov2004}
Novoselov,~K.~S.; Geim,~A.~K.; Morozov,~S.; Jiang,~D.; Zhang,~Y.; Dubonos,~S.;
  Grigorieva,~I.; Firsov,~A. Electric field effect in atomically thin carbon
  films. \emph{Science} \textbf{2004}, \emph{306}, 666--669\relax
\mciteBstWouldAddEndPuncttrue
\mciteSetBstMidEndSepPunct{\mcitedefaultmidpunct}
{\mcitedefaultendpunct}{\mcitedefaultseppunct}\relax
\EndOfBibitem
\bibitem[Wang et~al.(2012)Wang, Kalantar~Zadeh, Kis, Coleman, and
  Strano]{wangqh-natnano-2012}
Wang,~Q.~H.; Kalantar~Zadeh,~K.; Kis,~A.; Coleman,~J.~N.; Strano,~M.~S.
  Electronics and Optoelectronics of Two-Dimensional Transition Metal
  Dichalcogenides. \emph{Nat. Nanotechnol.} \textbf{2012}, \emph{7},
  699--712\relax
\mciteBstWouldAddEndPuncttrue
\mciteSetBstMidEndSepPunct{\mcitedefaultmidpunct}
{\mcitedefaultendpunct}{\mcitedefaultseppunct}\relax
\EndOfBibitem
\bibitem[Chhowalla et~al.(2013)Chhowalla, Shin, Eda, Li, Loh, and
  Zhang]{Chhowalla-natchem-2013}
Chhowalla,~M.; Shin,~H.~S.; Eda,~G.; Li,~L.~J.; Loh,~K.~P.; Zhang,~H. The
  Chemistry of Two-Dimensional Layered Transition Metal Dichalcogenide
  Nanosheets. \emph{Nat. Chem.} \textbf{2013}, \emph{5}, 263--275\relax
\mciteBstWouldAddEndPuncttrue
\mciteSetBstMidEndSepPunct{\mcitedefaultmidpunct}
{\mcitedefaultendpunct}{\mcitedefaultseppunct}\relax
\EndOfBibitem
\bibitem[Zhang et~al.(2015)Zhang, Qiao, Shi, Wu, Jiang, and
  Tan]{zhangx-csr-2015}
Zhang,~X.; Qiao,~X.~F.; Shi,~W.; Wu,~J.~B.; Jiang,~D.~S.; Tan,~P.~H. Phonon and
  Raman Scattering of Two-Dimensional Transition Metal Dichalcogenides from
  Monolayer, Multilayer to Bulk Material. \emph{Chem. Soc. Rev.} \textbf{2015},
  \emph{44}, 2757--2785\relax
\mciteBstWouldAddEndPuncttrue
\mciteSetBstMidEndSepPunct{\mcitedefaultmidpunct}
{\mcitedefaultendpunct}{\mcitedefaultseppunct}\relax
\EndOfBibitem
\bibitem[Qiao et~al.(2014)Qiao, Kong, Hu, Yang, and Ji]{qiaojs-natcom-2014}
Qiao,~J.~S.; Kong,~X.~H.; Hu,~Z.~X.; Yang,~F.; Ji,~W. High-Mobility Transport
  Anisotropy and Linear Dichroism in Few-Layer Black Phosphorus. \emph{Nat.
  Commun.} \textbf{2014}, \emph{5}, 4475\relax
\mciteBstWouldAddEndPuncttrue
\mciteSetBstMidEndSepPunct{\mcitedefaultmidpunct}
{\mcitedefaultendpunct}{\mcitedefaultseppunct}\relax
\EndOfBibitem
\bibitem[Liu et~al.(2014)Liu, Neal, Zhu, Luo, Xu, Tomanek, and
  Ye]{liuh-acsn-2014}
Liu,~H.; Neal,~A.~T.; Zhu,~Z.; Luo,~Z.; Xu,~X.~F.; Tomanek,~D.; Ye,~P.~D.
  Phosphorene: An Unexplored 2D Semiconductor with a High Hole Mobility.
  \emph{ACS Nano} \textbf{2014}, \emph{8}, 4033--4041\relax
\mciteBstWouldAddEndPuncttrue
\mciteSetBstMidEndSepPunct{\mcitedefaultmidpunct}
{\mcitedefaultendpunct}{\mcitedefaultseppunct}\relax
\EndOfBibitem
\bibitem[Li et~al.(2014)Li, Yu, Ye, Ge, Ou, Wu, Feng, Chen, and
  Zhang]{zhangyb-nnano-2014}
Li,~L.~K.; Yu,~Y.~J.; Ye,~G.~J.; Ge,~Q.~Q.; Ou,~X.~D.; Wu,~H.; Feng,~D.~L.;
  Chen,~X.~H.; Zhang,~Y.~B. Black Phosphorus Field-Effect Transistors.
  \emph{Nat. Nanotechnol.} \textbf{2014}, \emph{9}, 372--377\relax
\mciteBstWouldAddEndPuncttrue
\mciteSetBstMidEndSepPunct{\mcitedefaultmidpunct}
{\mcitedefaultendpunct}{\mcitedefaultseppunct}\relax
\EndOfBibitem
\bibitem[Wang et~al.(2015)Wang, Jones, Seyler, Tran, Jia, Zhao, Wang, Yang, Xu,
  and Xia]{wang2015highly}
Wang,~X.; Jones,~A.~M.; Seyler,~K.~L.; Tran,~V.; Jia,~Y.; Zhao,~H.; Wang,~H.;
  Yang,~L.; Xu,~X.; Xia,~F. Highly anisotropic and robust excitons in monolayer
  black phosphorus. \emph{Nat. Nanotechnol.} \textbf{2015}, \relax
\mciteBstWouldAddEndPunctfalse
\mciteSetBstMidEndSepPunct{\mcitedefaultmidpunct}
{}{\mcitedefaultseppunct}\relax
\EndOfBibitem
\bibitem[Zhao et~al.(2014)Zhao, Lo, Zhang, Sun, Tan, Uher, Wolverton, Dravid,
  and Kanatzidis]{nature-2014-SnSe}
Zhao,~L.-D.; Lo,~S.-H.; Zhang,~Y.; Sun,~H.; Tan,~G.; Uher,~C.; Wolverton,~C.;
  Dravid,~V.~P.; Kanatzidis,~M.~G. Ultralow thermal conductivity and high
  thermoelectric figure of merit in SnSe crystals. \emph{Nature} \textbf{2014},
  \emph{508}, 373--377\relax
\mciteBstWouldAddEndPuncttrue
\mciteSetBstMidEndSepPunct{\mcitedefaultmidpunct}
{\mcitedefaultendpunct}{\mcitedefaultseppunct}\relax
\EndOfBibitem
\bibitem[Zhao et~al.(2015)Zhao, Tan, Hao, He, Pei, Chi, Wang, Gong, Xu, Dravid,
  et~al. others]{Science-2015-SnSe}
Zhao,~L.-D.; Tan,~G.; Hao,~S.; He,~J.; Pei,~Y.; Chi,~H.; Wang,~H.; Gong,~S.;
  Xu,~H.; Dravid,~V.~P. et~al.  Ultrahigh power factor and thermoelectric
  performance in hole-doped single-crystal SnSe. \emph{Science} \textbf{2015},
  aad3749\relax
\mciteBstWouldAddEndPuncttrue
\mciteSetBstMidEndSepPunct{\mcitedefaultmidpunct}
{\mcitedefaultendpunct}{\mcitedefaultseppunct}\relax
\EndOfBibitem
\bibitem[Liu et~al.(2015)Liu, Fu, Wang, Feng, Liu, Wan, Zhou, Wang, Shao, Ho,
  et~al. others]{ReS2-2015-integrated}
Liu,~E.; Fu,~Y.; Wang,~Y.; Feng,~Y.; Liu,~H.; Wan,~X.; Zhou,~W.; Wang,~B.;
  Shao,~L.; Ho,~C.-H. et~al.  Integrated digital inverters based on
  two-dimensional anisotropic ReS$_2$ field-effect transistors. \emph{Nature
  Commun.} \textbf{2015}, \emph{6}, 6991\relax
\mciteBstWouldAddEndPuncttrue
\mciteSetBstMidEndSepPunct{\mcitedefaultmidpunct}
{\mcitedefaultendpunct}{\mcitedefaultseppunct}\relax
\EndOfBibitem
\bibitem[Zhao et~al.(2015)Zhao, Wu, Zhong, Guo, Wang, Xia, Yang, Tan, and
  Wang]{wangh-2015}
Zhao,~H.; Wu,~J.~B.; Zhong,~H.~X.; Guo,~Q.~S.; Wang,~X.~M.; Xia,~F.~N.;
  Yang,~L.; Tan,~P.~H.; Wang,~H. Interlayer Interactions in Anisotropic
  Atomically-thin Rhenium Diselenide. \emph{Nano Res.} \textbf{2015}, \emph{8},
  3651¨C3661\relax
\mciteBstWouldAddEndPuncttrue
\mciteSetBstMidEndSepPunct{\mcitedefaultmidpunct}
{\mcitedefaultendpunct}{\mcitedefaultseppunct}\relax
\EndOfBibitem
\bibitem[Mak et~al.(2010)Mak, Lee, Hone, Shan, and Heinz]{Heinz-prl-2010}
Mak,~K.~F.; Lee,~C.; Hone,~J.; Shan,~J.; Heinz,~T.~F. Atomically Thin MoS$_2$:
  A New Direct-Gap Semiconductor. \emph{Phys. Rev. Lett.} \textbf{2010},
  \emph{105}, 136805\relax
\mciteBstWouldAddEndPuncttrue
\mciteSetBstMidEndSepPunct{\mcitedefaultmidpunct}
{\mcitedefaultendpunct}{\mcitedefaultseppunct}\relax
\EndOfBibitem
\bibitem[Ling et~al.(2015)Ling, Liang, Huang, Puretzky, Geohegan, Sumpter,
  Kong, Meunier, and Dresselhaus]{ling2015low}
Ling,~X.; Liang,~L.; Huang,~S.; Puretzky,~A.~A.; Geohegan,~D.~B.;
  Sumpter,~B.~G.; Kong,~J.; Meunier,~V.; Dresselhaus,~M.~S. Low-frequency
  Interlayer Breathing Modes in Few-layer Black Phosphorus. \emph{Nano Lett.}
  \textbf{2015}, \emph{15}, 4080--4088\relax
\mciteBstWouldAddEndPuncttrue
\mciteSetBstMidEndSepPunct{\mcitedefaultmidpunct}
{\mcitedefaultendpunct}{\mcitedefaultseppunct}\relax
\EndOfBibitem
\bibitem[Jones et~al.(2013)Jones, Yu, Ghimire, Wu, Aivazian, Ross, Zhao, Yan,
  Mandrus, Xiao, et~al. others]{WSe2-nn-2013-optical}
Jones,~A.~M.; Yu,~H.; Ghimire,~N.~J.; Wu,~S.; Aivazian,~G.; Ross,~J.~S.;
  Zhao,~B.; Yan,~J.; Mandrus,~D.~G.; Xiao,~D. et~al.  Optical generation of
  excitonic valley coherence in monolayer WSe$_2$. \emph{Nature Nanotech.}
  \textbf{2013}, \emph{8}, 634--638\relax
\mciteBstWouldAddEndPuncttrue
\mciteSetBstMidEndSepPunct{\mcitedefaultmidpunct}
{\mcitedefaultendpunct}{\mcitedefaultseppunct}\relax
\EndOfBibitem
\bibitem[Tongay et~al.()Tongay, Sahin, Ko, Luce, Fan, Liu, Zhou, Huang, Ho,
  Yan, et~al. others]{ReS2-2014-monolayer}
Tongay,~S.; Sahin,~H.; Ko,~C.; Luce,~A.; Fan,~W.; Liu,~K.; Zhou,~J.;
  Huang,~Y.-S.; Ho,~C.-H.; Yan,~J. et~al.  Monolayer behaviour in bulk ReS2 due
  to electronic and vibrational decoupling. \emph{Nature Commun.} \emph{5},
  3252\relax
\mciteBstWouldAddEndPuncttrue
\mciteSetBstMidEndSepPunct{\mcitedefaultmidpunct}
{\mcitedefaultendpunct}{\mcitedefaultseppunct}\relax
\EndOfBibitem
\bibitem[Ali et~al.(2014)Ali, Xiong, Flynn, Tao, Gibson, Schoop, Liang,
  Haldolaarachchige, Hirschberger, Ong, et~al. others]{WTe2-2014-large}
Ali,~M.~N.; Xiong,~J.; Flynn,~S.; Tao,~J.; Gibson,~Q.~D.; Schoop,~L.~M.;
  Liang,~T.; Haldolaarachchige,~N.; Hirschberger,~M.; Ong,~N. et~al.  Large,
  non-saturating magnetoresistance in WTe$_2$. \emph{Nature} \textbf{2014},
  \emph{514}, 205--208\relax
\mciteBstWouldAddEndPuncttrue
\mciteSetBstMidEndSepPunct{\mcitedefaultmidpunct}
{\mcitedefaultendpunct}{\mcitedefaultseppunct}\relax
\EndOfBibitem
\bibitem[Pletikosi{\'c} et~al.(2014)Pletikosi{\'c}, Ali, Fedorov, Cava, and
  Valla]{prl-2014-electronic}
Pletikosi{\'c},~I.; Ali,~M.~N.; Fedorov,~A.; Cava,~R.; Valla,~T. Electronic
  Structure Basis for the Extraordinary Magnetoresistance in WTe$_2$.
  \emph{Phys. Rev. Lett.} \textbf{2014}, \emph{113}, 216601\relax
\mciteBstWouldAddEndPuncttrue
\mciteSetBstMidEndSepPunct{\mcitedefaultmidpunct}
{\mcitedefaultendpunct}{\mcitedefaultseppunct}\relax
\EndOfBibitem
\bibitem[Tan et~al.(2012)Tan, Han, Zhao, Wu, Chang, Wang, Wang, Bonini,
  Marzari, Pugno, and et. al.]{tanph-natm-2012}
Tan,~P.~H.; Han,~W.~P.; Zhao,~W.~J.; Wu,~Z.~H.; Chang,~K.; Wang,~H.;
  Wang,~Y.~F.; Bonini,~N.; Marzari,~N.; Pugno,~N. et~al.  The Shear Mode of
  Multilayer Graphene. \emph{Nat. Mater.} \textbf{2012}, \emph{11},
  294--300\relax
\mciteBstWouldAddEndPuncttrue
\mciteSetBstMidEndSepPunct{\mcitedefaultmidpunct}
{\mcitedefaultendpunct}{\mcitedefaultseppunct}\relax
\EndOfBibitem
\bibitem[Qiao et~al.(2015)Qiao, Li, Zhang, Shi, Wu, Chen, and
  Tan]{qiaoxf-apl-2015}
Qiao,~X.~F.; Li,~X.~L.; Zhang,~X.; Shi,~W.; Wu,~J.~B.; Chen,~T.; Tan,~P.~H.
  Substrate-Free Layer-Number Identification of Two-Dimensional Materials: A
  Case of Mo$_0.5$W$_0.5$S$_2$ Alloy. \emph{Appl. Phys. Lett.} \textbf{2015},
  \emph{106}, 223102\relax
\mciteBstWouldAddEndPuncttrue
\mciteSetBstMidEndSepPunct{\mcitedefaultmidpunct}
{\mcitedefaultendpunct}{\mcitedefaultseppunct}\relax
\EndOfBibitem
\bibitem[Lu et~al.(2015)Lu, Utama, Lin, Luo, Zhao, Zhang, Pantelides, Zhou,
  Quek, and Xiong]{lu2015rapid}
Lu,~X.; Utama,~M.; Lin,~J.; Luo,~X.; Zhao,~Y.; Zhang,~J.; Pantelides,~S.~T.;
  Zhou,~W.; Quek,~S.~Y.; Xiong,~Q. Rapid and Nondestructive Identification of
  Polytypism and Stacking Sequences in Few-Layer Molybdenum Diselenide by Raman
  Spectroscopy. \emph{Adv. Mater.} \textbf{2015}, \emph{27}, 4502--4508\relax
\mciteBstWouldAddEndPuncttrue
\mciteSetBstMidEndSepPunct{\mcitedefaultmidpunct}
{\mcitedefaultendpunct}{\mcitedefaultseppunct}\relax
\EndOfBibitem
\bibitem[Liu et~al.(2015)Liu, Du, Deng, and Peide]{BP-CSR-2015}
Liu,~H.; Du,~Y.; Deng,~Y.; Peide,~D.~Y. Semiconducting black phosphorus:
  synthesis, transport properties and electronic applications. \emph{Chem. Soc.
  Rev.} \textbf{2015}, \emph{44}, 2732--2743\relax
\mciteBstWouldAddEndPuncttrue
\mciteSetBstMidEndSepPunct{\mcitedefaultmidpunct}
{\mcitedefaultendpunct}{\mcitedefaultseppunct}\relax
\EndOfBibitem
\bibitem[Wu et~al.(2014)Wu, Zhang, Ij{\"a}s, Han, Qiao, Li, Jiang, Ferrari, and
  Tan]{wu-2014-resonant}
Wu,~J.-B.; Zhang,~X.; Ij{\"a}s,~M.; Han,~W.-P.; Qiao,~X.-F.; Li,~X.-L.;
  Jiang,~D.-S.; Ferrari,~A.~C.; Tan,~P.-H. Resonant Raman spectroscopy of
  twisted multilayer graphene. \emph{Nature Commun.} \textbf{2014}, \emph{5},
  5309\relax
\mciteBstWouldAddEndPuncttrue
\mciteSetBstMidEndSepPunct{\mcitedefaultmidpunct}
{\mcitedefaultendpunct}{\mcitedefaultseppunct}\relax
\EndOfBibitem
\bibitem[Puretzky et~al.(2015)Puretzky, Liang, Li, Xiao, Wang, Mahjouri-Samani,
  Basile, Idrobo, Sumpter, Meunier, et~al. others]{puretzky2015low}
Puretzky,~A.~A.; Liang,~L.; Li,~X.; Xiao,~K.; Wang,~K.; Mahjouri-Samani,~M.;
  Basile,~L.; Idrobo,~J.~C.; Sumpter,~B.~G.; Meunier,~V. et~al.  Low-Frequency
  Raman'Fingerprints' of Two-Dimensional Metal Dichalcogenide Layer Stacking
  Configurations. \emph{ACS nano} \textbf{2015}, \emph{9}, 6333--6342\relax
\mciteBstWouldAddEndPuncttrue
\mciteSetBstMidEndSepPunct{\mcitedefaultmidpunct}
{\mcitedefaultendpunct}{\mcitedefaultseppunct}\relax
\EndOfBibitem
\bibitem[Dos~Santos et~al.(2007)Dos~Santos, Peres, and Neto]{dos-2007-graphene}
Dos~Santos,~J.~L.; Peres,~N.; Neto,~A.~C. Graphene bilayer with a twist:
  Electronic structure. \emph{Phys. Rev. Lett.} \textbf{2007}, \emph{99},
  256802\relax
\mciteBstWouldAddEndPuncttrue
\mciteSetBstMidEndSepPunct{\mcitedefaultmidpunct}
{\mcitedefaultendpunct}{\mcitedefaultseppunct}\relax
\EndOfBibitem
\bibitem[Wu et~al.(2015)Wu, Hu, Zhang, Han, Lu, W., Qiao, Ijas, Milana, Ji,
  Ferrari, and Tan]{wujb-acsn-2015}
Wu,~J.~B.; Hu,~Z.~X.; Zhang,~X.; Han,~W.~P.; Lu,~Y.; W.,~S.; Qiao,~X.~F.;
  Ijas,~M.; Milana,~S.; Ji,~W. et~al.  Interface Coupling in Twisted Multilayer
  Graphene by Resonant Raman Spectroscopy of Layer Breathing Modes. \emph{ACS
  Nano} \textbf{2015}, \emph{9}, 7440--7449\relax
\mciteBstWouldAddEndPuncttrue
\mciteSetBstMidEndSepPunct{\mcitedefaultmidpunct}
{\mcitedefaultendpunct}{\mcitedefaultseppunct}\relax
\EndOfBibitem
\bibitem[Bonaccorso et~al.(2013)Bonaccorso, Tan, and
  Ferrari]{bonaccorso-2013-multiwall}
Bonaccorso,~F.; Tan,~P.-H.; Ferrari,~A.~C. Multiwall nanotubes, multilayers,
  and hybrid nanostructures: new frontiers for technology and Raman
  spectroscopy. \emph{ACS Nano} \textbf{2013}, \emph{7}, 1838--1844\relax
\mciteBstWouldAddEndPuncttrue
\mciteSetBstMidEndSepPunct{\mcitedefaultmidpunct}
{\mcitedefaultendpunct}{\mcitedefaultseppunct}\relax
\EndOfBibitem
\bibitem[Feng et~al.(2015)Feng, Zhou, Wang, Zhou, Liu, Fu, Ni, Wu, Yuan, Miao,
  Wang, Wan, and Xing]{PhysRevB.92.054110}
Feng,~Y.; Zhou,~W.; Wang,~Y.; Zhou,~J.; Liu,~E.; Fu,~Y.; Ni,~Z.; Wu,~X.;
  Yuan,~H.; Miao,~F. et~al.  Raman vibrational spectra of bulk to monolayer
  $\mathrm{Re}{\mathrm{S}}_{2}$ with lower symmetry. \emph{Phys. Rev. B}
  \textbf{2015}, \emph{92}, 054110\relax
\mciteBstWouldAddEndPuncttrue
\mciteSetBstMidEndSepPunct{\mcitedefaultmidpunct}
{\mcitedefaultendpunct}{\mcitedefaultseppunct}\relax
\EndOfBibitem
\bibitem[Lamfers et~al.(1996)Lamfers, Meetsma, Wiegers, and
  De~Boer]{lamfers1996crystal}
Lamfers,~H.-J.; Meetsma,~A.; Wiegers,~G.; De~Boer,~J. The crystal structure of
  some rhenium and technetium dichalcogenides. \emph{J. Alloy. Compd.}
  \textbf{1996}, \emph{241}, 34--39\relax
\mciteBstWouldAddEndPuncttrue
\mciteSetBstMidEndSepPunct{\mcitedefaultmidpunct}
{\mcitedefaultendpunct}{\mcitedefaultseppunct}\relax
\EndOfBibitem
\bibitem[Kertesz and Hoffmann(1984)Kertesz, and
  Hoffmann]{kertesz1984octahedral}
Kertesz,~M.; Hoffmann,~R. Octahedral vs. trigonal-prismatic coordination and
  clustering in transition-metal dichalcogenides. \emph{J. Am. Chem. Soc.}
  \textbf{1984}, \emph{106}, 3453--3460\relax
\mciteBstWouldAddEndPuncttrue
\mciteSetBstMidEndSepPunct{\mcitedefaultmidpunct}
{\mcitedefaultendpunct}{\mcitedefaultseppunct}\relax
\EndOfBibitem
\bibitem[Fang et~al.(1997)Fang, Wiegers, Haas, and
  De~Groot]{fang1997electronic}
Fang,~C.; Wiegers,~G.; Haas,~C.; De~Groot,~R. Electronic structures of, and in
  the real and the hypothetical undistorted structures. \emph{J. Phys.:
  Condens. Matter.} \textbf{1997}, \emph{9}, 4411\relax
\mciteBstWouldAddEndPuncttrue
\mciteSetBstMidEndSepPunct{\mcitedefaultmidpunct}
{\mcitedefaultendpunct}{\mcitedefaultseppunct}\relax
\EndOfBibitem
\bibitem[Nagler et~al.(2015)Nagler, Plechinger, Sch{\"u}ller, and
  Korn]{Nagler-2015-C}
Nagler,~P.; Plechinger,~G.; Sch{\"u}ller,~C.; Korn,~T. Observation of
  anisotropic interlayer Raman modes in few-layer ReS2. \emph{physica status
  solidi (RRL)-Rapid Research Letters} \textbf{2015}, \emph{9999}\relax
\mciteBstWouldAddEndPuncttrue
\mciteSetBstMidEndSepPunct{\mcitedefaultmidpunct}
{\mcitedefaultendpunct}{\mcitedefaultseppunct}\relax
\EndOfBibitem
\bibitem[Zhang et~al.(2013)Zhang, Han, Wu, Milana, Lu, Li, Ferrari, and
  Tan]{zhangx-prb-2013}
Zhang,~X.; Han,~W.~P.; Wu,~J.~B.; Milana,~S.; Lu,~Y.; Li,~Q.~Q.;
  Ferrari,~A.~C.; Tan,~P.~H. Raman Spectroscopy of Shear and Layer Breathing
  Modes in Multilayer MoS$_2$. \emph{Phys. Rev. B} \textbf{2013}, \emph{87},
  115413\relax
\mciteBstWouldAddEndPuncttrue
\mciteSetBstMidEndSepPunct{\mcitedefaultmidpunct}
{\mcitedefaultendpunct}{\mcitedefaultseppunct}\relax
\EndOfBibitem
\bibitem[Zhang et~al.(2016)Zhang, Han, Qiao, Tan, Wang, Zhang, and
  Tan]{zhangxin-Carbon-2016}
Zhang,~X.; Han,~W.-P.; Qiao,~X.-F.; Tan,~Q.-H.; Wang,~Y.-F.; Zhang,~J.;
  Tan,~P.-H. Raman characterization of AB-and ABC-stacked few-layer graphene by
  interlayer shear modes. \emph{Carbon} \textbf{2016}, 118--122\relax
\mciteBstWouldAddEndPuncttrue
\mciteSetBstMidEndSepPunct{\mcitedefaultmidpunct}
{\mcitedefaultendpunct}{\mcitedefaultseppunct}\relax
\EndOfBibitem
\bibitem[Lui et~al.(2015)Lui, Ye, Ji, Chiu, Chou, Andersen, Means-Shively,
  Anderson, Wu, Kidd, Lee, and He]{tBL-MoS2}
Lui,~C.~H.; Ye,~Z.~P.; Ji,~C.; Chiu,~K.~C.; Chou,~C.~T.; Andersen,~T.~I.;
  Means-Shively,~C.; Anderson,~H.; Wu,~J.~M.; Kidd,~T. et~al.  Observation of
  Interlayer Phonon Modes in Van der Waals Heterostructures. \emph{Phys. Rev.
  B} \textbf{2015}, \emph{91}, 165403\relax
\mciteBstWouldAddEndPuncttrue
\mciteSetBstMidEndSepPunct{\mcitedefaultmidpunct}
{\mcitedefaultendpunct}{\mcitedefaultseppunct}\relax
\EndOfBibitem
\bibitem[Jiang et~al.(2014)Jiang, Liu, Huang, Zhang, Li, Gong, Shen, Liu, and
  Wu]{jiang2014valley}
Jiang,~T.; Liu,~H.; Huang,~D.; Zhang,~S.; Li,~Y.; Gong,~X.; Shen,~Y.-R.;
  Liu,~W.-T.; Wu,~S. Valley and band structure engineering of folded MoS2
  bilayers. \emph{Nat. Nanotechnol.} \textbf{2014}, \emph{9}, 825--829\relax
\mciteBstWouldAddEndPuncttrue
\mciteSetBstMidEndSepPunct{\mcitedefaultmidpunct}
{\mcitedefaultendpunct}{\mcitedefaultseppunct}\relax
\EndOfBibitem
\bibitem[Zhao et~al.(2013)Zhao, Luo, Li, Zhang, Araujo, Gan, Wu, Zhang, Quek,
  Dresselhaus, et~al. others]{zhao2013interlayer}
Zhao,~Y.; Luo,~X.; Li,~H.; Zhang,~J.; Araujo,~P.~T.; Gan,~C.~K.; Wu,~J.;
  Zhang,~H.; Quek,~S.~Y.; Dresselhaus,~M.~S. et~al.  Interlayer breathing and
  shear modes in few-trilayer MoS2 and WSe2. \emph{Nano Lett.} \textbf{2013},
  \emph{13}, 1007--1015\relax
\mciteBstWouldAddEndPuncttrue
\mciteSetBstMidEndSepPunct{\mcitedefaultmidpunct}
{\mcitedefaultendpunct}{\mcitedefaultseppunct}\relax
\EndOfBibitem
\bibitem[Kaneta et~al.(1986)Kaneta, Katayama-Yoshida, and
  Morita]{kaneta1986lattice}
Kaneta,~C.; Katayama-Yoshida,~H.; Morita,~A. Lattice Dynamics of Black
  Phosphorus. I. Valence Force Field Model. \emph{Journal of the Physical
  Society of Japan} \textbf{1986}, \emph{55}, 1213--1223\relax
\mciteBstWouldAddEndPuncttrue
\mciteSetBstMidEndSepPunct{\mcitedefaultmidpunct}
{\mcitedefaultendpunct}{\mcitedefaultseppunct}\relax
\EndOfBibitem
\bibitem[Molina-Sanchez and Wirtz(2011)Molina-Sanchez, and
  Wirtz]{molina2011phonons}
Molina-Sanchez,~A.; Wirtz,~L. Phonons in single-layer and few-layer MoS$_2$ and
  WS$_2$. \emph{Phys. Rev. B} \textbf{2011}, \emph{84}, 155413\relax
\mciteBstWouldAddEndPuncttrue
\mciteSetBstMidEndSepPunct{\mcitedefaultmidpunct}
{\mcitedefaultendpunct}{\mcitedefaultseppunct}\relax
\EndOfBibitem
\bibitem[Blochl(1994)]{Blochl-prb-1994}
Blochl,~P.~E. Projector Augmented-Wave Method. \emph{Phys. Rev. B}
  \textbf{1994}, \emph{50}, 17953\relax
\mciteBstWouldAddEndPuncttrue
\mciteSetBstMidEndSepPunct{\mcitedefaultmidpunct}
{\mcitedefaultendpunct}{\mcitedefaultseppunct}\relax
\EndOfBibitem
\bibitem[Kresse and Joubert(1999)Kresse, and Joubert]{Kresse-prb-1999}
Kresse,~G.; Joubert,~D. From Ultrasoft Pseudopotentials to the Projector
  Augmented-Wave Method. \emph{Phys. Rev. B} \textbf{1999}, \emph{59},
  1758\relax
\mciteBstWouldAddEndPuncttrue
\mciteSetBstMidEndSepPunct{\mcitedefaultmidpunct}
{\mcitedefaultendpunct}{\mcitedefaultseppunct}\relax
\EndOfBibitem
\bibitem[Kresse and Furthmuller(1996)Kresse, and Furthmuller]{Kresse-prb-1996}
Kresse,~G.; Furthmuller,~J. Efficient Iterative Schemes for \textit{ab initio}
  Total-Energy Calculations using a Plane-Wave Basis Set. \emph{Phys. Rev. B}
  \textbf{1996}, \emph{54}, 11169\relax
\mciteBstWouldAddEndPuncttrue
\mciteSetBstMidEndSepPunct{\mcitedefaultmidpunct}
{\mcitedefaultendpunct}{\mcitedefaultseppunct}\relax
\EndOfBibitem
\bibitem[Baroni et~al.(2001)Baroni, de~Gironcoli, Dal~Corso, and
  Giannozzi]{Baroni-Rmp-2001}
Baroni,~S.; de~Gironcoli,~S.; Dal~Corso,~A.; Giannozzi,~P. Phonons and Related
  Crystal Properties from Density-Functional Perturbation Theory. \emph{Rev.
  Mod. Phys.} \textbf{2001}, \emph{73}, 515--562\relax
\mciteBstWouldAddEndPuncttrue
\mciteSetBstMidEndSepPunct{\mcitedefaultmidpunct}
{\mcitedefaultendpunct}{\mcitedefaultseppunct}\relax
\EndOfBibitem
\bibitem[Dion et~al.(2004)Dion, Rydberg, Schr\"oder, Langreth, and
  Lundqvist]{Dion-prl-2004}
Dion,~M.; Rydberg,~H.; Schr\"oder,~E.; Langreth,~D.~C.; Lundqvist,~B.~I. Van
  der Waals Density Functional for General Geometries. \emph{Phys. Rev. Lett.}
  \textbf{2004}, \emph{92}, 246401\relax
\mciteBstWouldAddEndPuncttrue
\mciteSetBstMidEndSepPunct{\mcitedefaultmidpunct}
{\mcitedefaultendpunct}{\mcitedefaultseppunct}\relax
\EndOfBibitem
\bibitem[Lee et~al.(2010)Lee, Murray, Kong, Lundqvist, and
  Langreth]{LeeK-prb-2010}
Lee,~K.; Murray,~E.~D.; Kong,~L.~Z.; Lundqvist,~B.~I.; Langreth,~D.~C.
  Higher-Accuracy Van der Waals Density Functional. \emph{Phys. Rev. B}
  \textbf{2010}, \emph{82}, 081101\relax
\mciteBstWouldAddEndPuncttrue
\mciteSetBstMidEndSepPunct{\mcitedefaultmidpunct}
{\mcitedefaultendpunct}{\mcitedefaultseppunct}\relax
\EndOfBibitem
\bibitem[Klimes et~al.(2010)Klimes, Bowler, and Michaelides]{Klimes-jp-2010}
Klimes,~J.; Bowler,~D.~R.; Michaelides,~A. Chemical Accuracy for the Van der
  Waals Density Functional. \emph{Journal of Physics: Condensed Matter}
  \textbf{2010}, \emph{22}, 022201\relax
\mciteBstWouldAddEndPuncttrue
\mciteSetBstMidEndSepPunct{\mcitedefaultmidpunct}
{\mcitedefaultendpunct}{\mcitedefaultseppunct}\relax
\EndOfBibitem
\bibitem[Klimes et~al.(2011)Klimes, Bowler, and Michaelides]{Klimes-prb-2011}
Klimes,~J.; Bowler,~D.~R.; Michaelides,~A. Van der Waals Density Functionals
  Applied to Solids. \emph{Phys. Rev. B} \textbf{2011}, \emph{83}, 195131\relax
\mciteBstWouldAddEndPuncttrue
\mciteSetBstMidEndSepPunct{\mcitedefaultmidpunct}
{\mcitedefaultendpunct}{\mcitedefaultseppunct}\relax
\EndOfBibitem
\bibitem[Hu et~al.(2014)Hu, Lan, and Ji]{huzx-sp-2014}
Hu,~Z.~X.; Lan,~H.~P.; Ji,~W. Role of the Dispersion Force in Modeling the
  Interfacial Properties of Molecule-Metal Interfaces: Adsorption of Thiophene
  on Copper Surfaces. \emph{Scientific Reports} \textbf{2014}, \emph{4},
  5036\relax
\mciteBstWouldAddEndPuncttrue
\mciteSetBstMidEndSepPunct{\mcitedefaultmidpunct}
{\mcitedefaultendpunct}{\mcitedefaultseppunct}\relax
\EndOfBibitem
\bibitem[Hong et~al.(2015)Hong, Hu, Probert, Li, Lv, Yang, Gu, Mao, Feng, Xie,
  Zhang, Wu, Zhang, Jin, Ji, Zhang, Yuan, and Zhang]{hongjh-natcommu-2015}
Hong,~J.~H.; Hu,~Z.~X.; Probert,~M.; Li,~K.; Lv,~D.; Yang,~X.~N.; Gu,~L.;
  Mao,~N.~N.; Feng,~Q.~L.; Xie,~L.~M. et~al.  Exploring Atomic Defects in
  Molybdenum Disulphide Monolayers. \emph{Nat. Commun.} \textbf{2015},
  \emph{6}, 6293\relax
\mciteBstWouldAddEndPuncttrue
\mciteSetBstMidEndSepPunct{\mcitedefaultmidpunct}
{\mcitedefaultendpunct}{\mcitedefaultseppunct}\relax
\EndOfBibitem
\bibitem[Hu et~al.(2015)Hu, Kong, Qiao, Normand, and Ji]{huzx-ns-2015}
Hu,~Z.-X.; Kong,~X.; Qiao,~J.; Normand,~B.; Ji,~W. Interlayer electronic
  hybridization leads to exceptional thickness-dependent vibrational properties
  in few-layer black phosphorus. \emph{arXiv preprint arXiv:1503.06735}
  \textbf{2015}, \relax
\mciteBstWouldAddEndPunctfalse
\mciteSetBstMidEndSepPunct{\mcitedefaultmidpunct}
{}{\mcitedefaultseppunct}\relax
\EndOfBibitem
\bibitem[He et~al.(2015)He, Yan, Yin, Ye, Ye, Cheng, Li, and
  Lui]{he2015coupling}
He,~R.; Yan,~J.-A.; Yin,~Z.; Ye,~Z.; Ye,~G.; Cheng,~J.; Li,~J.; Lui,~C.
  Coupling and stacking order of ReS2 atomic layers revealed by
  ultralow-frequency Raman spectroscopy. \emph{arXiv preprint arXiv:1512.00092}
  \textbf{2015}, \relax
\mciteBstWouldAddEndPunctfalse
\mciteSetBstMidEndSepPunct{\mcitedefaultmidpunct}
{}{\mcitedefaultseppunct}\relax
\EndOfBibitem
\bibitem[Lorchat et~al.(2015)Lorchat, Froehlicher, and
  Berciaud]{lorchat2015splitting}
Lorchat,~E.; Froehlicher,~G.; Berciaud,~S. Splitting of interlayer shear modes
  and photon energy dependent anisotropic Raman response in $ N $-layer ReSe $
  \_2 $ and ReS $ \_2$. \emph{arXiv preprint arXiv:1512.03842} \textbf{2015},
  \relax
\mciteBstWouldAddEndPunctfalse
\mciteSetBstMidEndSepPunct{\mcitedefaultmidpunct}
{}{\mcitedefaultseppunct}\relax
\EndOfBibitem
\end{mcitethebibliography}
\end{document}